\documentclass[epjST]{svjour}

\usepackage{graphics}
\usepackage{url}
\begin{document}
\title{Challenges in Complex Systems Science}

\author{Maxi San Miguel\inst{1}\fnmsep\thanks{\email{maxi@ifisc.uib-csic.es}}
\and Jeffrey H. Johnson\inst{2}
\and Janos Kertesz\inst{3}
\and Kimmo Kaski\inst{4}
\and Albert D{\'\i}az-Guilera\inst{5}
\and Robert S. MacKay\inst{6}
\and Vittorio Loreto\inst{7,8}
\and P\'eter \'Erdi\inst{9,10}
\and Dirk Helbing\inst{11}
}
\small
\institute{
$^1$IFISC (CSIC-UIB), Campus Universitat Illes Balears, E-07071 Palma de Mallorca, Spain\\
$^2$Faculty of Mathematics, Computing \& Technology, The Open University, MK7 6AA, UK\\
$^3$Institute of Physics, Budapest Univ. of Technology \& Economics, Budafoki \'ut 8, Budapest\\
$^4$Dept. of Biomedical Engineering \& Computational Science, FI-00076 Aalto, Finland\\
$^5$Dept. Fisica Fonamental,  Universitat de Barcelona, E-08028 Barcelona, Spain\\
$^6$Mathematics Institute \& Centre for Complexity Science,
University of Warwick, CV4 7AL\\
$^{7}$Sapienza Universty of Rome, Physics Dept., Rome, Italy\\
$^{8}$ISI Foundation, Turin, Italy\\
$^9$Institute for Particle and Nuclear Physics, Wigner Research Centre for Physics, Hungarian Academy of Sciences, Budapest \\
$^{10}$Center for Complex Systems Studies, Kalamazoo College, Michigan, MI 49006, USA\\
$^{11}$ETH Z\"{u}rich, Clausiusstrasse 50, 8092 Z\"{u}rich, Switzerland\\
}
\normalsize

\abstract{
FuturICT foundations are social science, complex systems science, and ICT.
The main concerns and challenges in the science of complex systems in the context of FuturICT are laid out in this paper with special emphasis on the {\it Complex Systems route to Social Sciences}. This include complex systems having: many heterogeneous interacting parts; multiple scales; complicated transition laws; unexpected or unpredicted emergence; sensitive dependence on initial conditions; path-dependent dynamics; networked hierarchical connectivities; interaction of autonomous agents; self-organisation; non-equilibrium dynamics; combinatorial explosion; adaptivity to changing environments; co-evolving subsystems; ill-defined boundaries; and multilevel dynamics. In this context, science is seen as the process of abstracting the dynamics of systems from data. This presents many challenges including: data gathering by large-scale experiment, participatory sensing and social computation, managing huge distributed dynamic and heterogeneous databases; moving from data to dynamical models, going beyond correlations to cause-effect relationships, understanding the relationship between simple and comprehensive models with appropriate choices of variables, ensemble modeling and data assimilation, modeling systems of systems of systems with many levels between micro and macro; and formulating new approaches to prediction, forecasting, and risk, especially in systems that can reflect on and change their behaviour in response to predictions, and systems whose apparently predictable behaviour is disrupted by apparently unpredictable rare or extreme events. These challenges are part of the FuturICT agenda.
}

%
\maketitle
\section{Introduction}
\label{intro}

Simplicity and sparsity of scientific description have always been regarded as a theoretical virtue. Aristotle in the {\it Posterior Analytics} says: ``We may assume the superiority {\it ceteris paribus} of the demonstration which derives from fewer postulates or hypotheses."  According to Newton ``Nature is pleased with simplicity, and affects not the pomp of superfluous causes", while we learn from Einstein that ``The grand aim of all science ... is to cover the greatest possible number of empirical facts by logical deductions from the smallest possible number of hypotheses or axioms."

Modern science started with the analysis of the simplest phenomena and physics, a reductionist science {\it par excellence}, emerged as the leading example of how the human mind can make sense of the apparent chaos of the phenomena surrounding us. The key to the early success of physics was that it studied objects that could be described in terms of a few variables, could be well separated from their environment, with well-targeted reproducible experiments that could be performed on them.  During the course of its development physics has learned how to tackle problems that are immensely more complicated than the free fall of balls from the Tower of Pisa, but the reductionist program remains one of its core motivations. The dream of a ``theory of everything" drives the quest for the ultimate building blocks of the Universe and for the explanation of its origin - an endeavor constituting one of the frontiers of science.

However, as stated by P. W. Anderson in 1972 \cite{Anderson} the reductionist hypothesis does not by any means imply a ``constructionist" hypothesis: the constructionist hypothesis breaks down when confronted with the twin difficulties of scale and complexity.  Most of the objects of scientific inquiry share these difficulties.  For example, a living being cannot be described in terms of a few variables, a human being cannot be separated from the rest of society without altering its nature fundamentally, and the functionality of our brain emerges from the network of interacting neurons.  These are examples of what nowadays are called complex systems.  A growing body of knowledge is being accumulated about these complex systems, a large number of groups are striving for a deeper understanding of their common features and an ever richer set of concepts and tools are being devised to tackle them. These developments are gradually leading up to what we believe is becoming a coherent and fundamental science of complexity \cite{pietronero_2008,HelbingComplexity}. Understanding the basic principles of complexity and emergent phenomena in complex systems is the other frontier of present day science. If the goal of particle physics is the ultimate analysis, that of complexity science is the ultimate synthesis. To promote this synthesis is one of the main motivations of the FuturICT program.

Other motivation stems from the ever-increasing relevance of complex-systems oriented approaches to social science \cite{pietronero_2010}. It may be surprising but the idea of a physical modeling of social phenomena \cite{rmp_2009} is in some sense older than the idea of statistical modeling of physical phenomena. The discovery of quantitative laws in the collective properties of a large number of people, as revealed, for example, by birth and death rates or crime statistics, was one of the catalysts in the development of statistics, and it led many scientists and philosophers to call for some quantitative understanding of how such precise regularities arise out of the apparently erratic behavior of single individuals. Hobbes, Laplace, Comte, Stuart Mill, and many others shared, to a different extent, this line of thought \cite{Ball_2004}. This point of view was well known to Maxwell and Boltzmann and probably played a role when they abandoned the idea of describing the trajectory of single particles and introduced a statistical description for gases, laying the foundations of modern statistical physics. The value of statistical laws for social science was foreseen also by Majorana \cite{Majorana_1942}.  But it is only in the past few years that the idea of approaching society within the framework of statistical physics has transformed from a philosophical declaration of principles to a concrete research effort involving a critical mass of scientists. The availability of new large databases as well as the appearance of brand new social phenomena, mostly related to the Information and Communication Technologies, and the tendency of social scientists to move toward the formulation of simplified models \cite{Schelling,Axelrod} and their quantitative analysis \cite{Centola}, have been instrumental in this change.  Nowadays the understanding of the dynamics of human societies and finding viable solutions to the enormous problems they are facing is a matter not only of knowledge, but survival. Demographic changes, migrations, destruction of the environment, depletion of resources, the structural instability of our economic and social systems are only some of the most prominent among these problems. FuturICT is devoted to the analysis and modeling of these complex and interwoven processes.

\section{Complex Systems}

Conventional wisdom suggests that simple systems behave simply, complex behavior arises from complex causes and that different systems behave differently. There is ample evidence, even in the physical sciences, that these statements are unfounded and not generally correct.  The roots of such oversimplified views are closely related to the problems of complex systems themselves, {\em e.g.} the choice of appropriate variables, the lack or multitude of scales, multilevel structure in both space and time, the level of description, and so on. It can be highly non-trivial to find the simple causes behind complex behavior and to select appropriate variables by which the possible generic behavior of the system would become apparent.

An important source of difficulties stems from the fundamental problem of the level of detail and complexity needed for the understanding of the structure, function, and response of a complex system of interest.  This is perhaps best gained with the analysis of data from well-devised experiments or from various datasets and proper modeling.  The success of this two-step empirical approach has to be judged in relation to the goal and purpose of the study process, while aiming for understanding, predicting\footnote
{One of the challenges for complex systems science is to better understand the term `prediction' and the part it plays in science and its applications. In this paper it will be used generally to include various different ways for describing the future behaviour of systems. Traditionally science has made {\em point predictions} that a system will be in a particular state at a particular point in future time. Beyond relatively short horizons, point predictions of systems that are sensitive to initial conditions become increasingly error prone, {e.g.} the weather. This is so
even when the underlying model is prefect, due to inevitable errors in measuring initial conditions. Some predictions are probabilistic based on estimates of relative frequencies, but this approach is inappropriate for `rare' events of measure zero. Predictions of social systems may be self-contradictory as the system reflects on and changes behaviour in response to the prediction, or self-fulfilling prophesies when policy forces events to happen.},
managing and even controlling the behavior of the system. In all this it should be emphasized that modeling goes hand in hand with the availability and use of the data. This is because the need for empirical data is paramount not only for the understanding and exploration of the features and phenomena of the systems of interest but also for calibrating and validating the models for improved usefulness in predicting, forecasting, and managing the behavior of the system.

Calibration can be a delicate task and much attention should be paid to it. Given the level of complexity and the related dimensionality of the problem the amount of data used for calibration should be chosen such that overfitting can be avoided in order to exclude spurious dependencies. On the other hand - as follows from the very nature of complex systems - as much data as possible should be used. This issue reflects the deep problem of computer science of finding the balance between avoiding both overfitting and oversimplification. Of course, it is also related to the  problem of finding the right variables mentioned previously.

In the new ICT-based frame of Social Sciences, the main problem is often not data availability but the challenge of extracting relevant knowledge from observational data and in devising useful data acquisition for answering specific questions of the behavior of the system of interest.

In relation to these specific questions we see an increasing role of more question-driven research, where massive, ICT-based data are already collected with a specific aim and even experiments are being devised (see Sect. 7.2). Such a trend could lead to or open new frontiers in studies of ICT-related social systems. To summarize we envisage that the study of a complex system should proceed in the following steps in which the use of data and models is essential:

\begin{itemize}

\item[i)] Exploration, observation and basic data acquisition,

\item[ii)] Identification of correlations, patterns, and mechanisms,

\item[iii)] Modeling,

\item[iv)] Model validation, implementation and prediction,

\item[v)] Construction of a theory.

\end{itemize}

In devising a model for a given system with a large number of constituents it is useful first to discuss how one would characterize the basic constituents and/or governing laws (if known). In this context a distinction can be made between \emph{complicated} systems and \emph{complex} systems.  Complicated systems are viewed to have a large number of components which behave in a well-understood way and have well-defined roles leading to the resulting effect, {\it e.g.} modern airplanes with millions of physical parts and even tens of millions of lines of software. Complex systems typically have a large number of components, where the interactions (however simple they may be on the individual level) lead to collective \emph{emergent} behaviours that cannot, even qualitatively, be derived as a plain resultant from the individual components' behavior. Paramount examples of complex systems are our brain and our societies.

All domain-based sciences such as physics, chemistry, biology, psychology, sociology, economics, robotics, medicine and business investigate systems that are complex in one way or another. These sciences investigate their domains in depth, which contrasts with the emerging science of complex systems which intersects the domains horizontally. By looking across the disciplines the methodology of complex systems provides two new perspectives: the first is that apparently different systems may have common properties and knowledge from one discipline can usefully feed into another; the second is that the science of complex systems is trans-disciplinary and it is creating new methods to combine the dynamical theories of many interacting social and technical subsystems.
Unlike domain-based sciences such as those mentioned above, complex
systems science is {\em integrative} - a science of systems of systems across many domains.

There is no agreement on what should be the precise definition of {\em complex} and there are many reasons as to why a system might be considered complex. Table \ref{table_1} reports a list of features typical of complex systems along with concrete examples of systems displaying those features. Of course many systems could exhibit several of these features. Any one of them can make systems appear complex, but together they can make systems very difficult to understand and control \cite{johnson}. A key characteristics of complex systems is their ability to reconfigure themselves to create new systems with completely different properties.

\begin{table}
  \label{table_1}
  \begin{tabular}{lllr|}
    \hline\hline\\
    many heterogeneous interacting parts &  cities, companies, climate, crowds\\
    & political parties, ecosystems                  \\\hline
    complicated transition laws &  markets, disease transmission, cascading failure\\
    & rioting, professional training              \\\hline
    unexpected or unpredictable emergence &  chemical systems, accidents, system breakdown\\
    & spontaneous social initiatives, foot and mouth disease                 \\\hline
    sensitive dependence on initial conditions &  weather systems, investments\\
    & traffic jams, forest fires              \\\hline
    path-dependent dynamics &  the evolution of the {\em qwerty} keyboard,\\
    & racial conflicts, first to market &\\\hline
    networked hierarchical connectivities &  social networks, ecosystems ,the Internet\\
    & voting systems, postal systems                        \\\hline
    interactions of autonomous agents &  road traffic, dinner parties\\
    & housing markets, soccer games, crowd dynamics                               \\\hline
    self-organisation or collective shifts &  revolutions, fashions, choirs\\
    & demonstrations, property rental markets                    \\\hline
    non-equilibrium dynamics & fighter aircraft, share prices, the weather\\
    & armed conflict, social networking      \\\hline
    combinatorial explosion &  chess, communications systems, \\
    & data states for a computer program &\\\hline
    adaptivity to changing environments &  biological systems,manufacturing design\\
    & retail systems, rebranding \\ \hline
    co-evolving subsystems &  land-use, transportation\\
    & computer virus software\\ \hline
    ill-defined boundaries &  genetically modified crops, nations,\\
    & pollution,  terrorism, markets\\\hline
    multilevel dynamics &  companies, armies, governments \\
    & aircraft, Internet, transportation \\
    \hline\hline
  \end{tabular}
  \caption{Reasons why systems might be considered to be complex}
\end{table}

Complex systems such as cities, the human body, or economies have dynamics at many different scales.  The presence of many scales or, even worse, the confluence of scales and lack of a characteristic scale that would allow the breakdown of the problem into sub-problems makes a standard reductionist micro-macro approach difficult.  This leads to the appearance of fat tails and self-similar distributions, {\em e.g.} Pareto-distributions of wealth, company size, capitalization, {\em etc.}.

Socio-technical systems have strong interactions leading to collective behaviour, building up macroscopic structures that act as top-down constraints on the microscopic degrees of freedom (mode slaving), {\em e.g.} long wavelength spin waves acting as an external field on the individual spins, or institutions, conventions, traditions, culture, {\em etc.} acting on, and largely conditioning, the agents that created them.  Thus ``the whole is more (or less) than the sum of its parts": {\em e.g.} cutting a horse in two does not result in two small horses; the merger of two successful companies (or universities) rarely creates a better company (but may create a monopoly); and uniting disparate nations into Yugoslavia led to disaster 70 years later.
Underlying the formation of wholes is the emergence of strong, long-range interactions and correlations in complex systems, that link distant parts. Complex systems are likely to feature non-local interactions in space and time.
This property often makes systems sensitive not only to initial conditions, but also to boundary conditions and small changes in the control parameters.

This is tightly related to ``irreducibility", {\em i.e.,} the impossibility of describing a complex system in terms of a few variables. (The local susceptibility is the sum of correlations measured from the given local element: if correlations are long ranged and the system is heterogeneous, the local susceptibility depends on a large number of variables.) Multi-attractor structure and the resulting path-dependence are related aspects. Complex systems may sometimes evolve slowly, but they are never in equilibrium (unless dead). For example: it is hard to predict biological evolution, but it may be possible to rationalize backwards.

Beyond a certain level of complexity systems not only reflect their prior evolution, but also start to learn, and modify their behaviour according to changes in the environment, conditions, {\em etc}.  At even higher levels they self-reflect, react to what they ``think" about themselves, or what is thought about them, {\em e.g.} self-fulfilling prophecies, collective myths, {\em etc.}.
For example, tourism may cause prices to rise as traders see the opportunities, but make local people resentful
with the self-perception of being second-class citizens.

Concerning controlling or regulating complex systems, the Law of Unintended Consequences has a pervasive effect. As a consequence of the irreducibility of complex systems, they cannot
always
be reliably regulated or controlled, but since social arrangements, markets, finance, {\em etc.} are man-made, one can strive to reduce their complexity to bring them into controllable regions. For example, it may be that some financial procedures and products create systems that are inherently volatile and unpredictable exposing society to risk of highly damaging outcomes. In these cases the contribution of science is to inform the regulators that this particular system, due to its complexity and unknown to its designers, is inherently unpredictable and undesirable. In this case the danger can mediated by {\em simplifying} the products and procedures to make them more predictable and controllable.

In the context of economic systems, neither central planning nor self-regulation seem to work, but biological regulation (an intricate network of positive and negative feedbacks, checks and balances on every level) appears to be capable of keeping a complex system in homeostasis, at least for an extended period.

\section{Open fundamental questions in Complexity Science}
\label{sec:1}
\noindent
\textbf{Simple versus comprehensive models.}

\vspace{0.08in}

Complex systems need not be complicated, but in real life they often are. Simple models are essential to uncover the basic mechanisms and provide insight into fundamental questions. However, in order to be able to predict
self-fulfilling prophecies, collective myths, {\em etc.}.
or forecast
the behavior of real systems one often has to go to more detailed, multi-parameter models. Both approaches have their justification and they are complementary. However, this kind of ``pluralistic" modeling \cite{pluralistic} does not mean the acceptance of different scientific truths, rather it could give more comprehensive perspective to the behavior of the system of interest. The models should be hierarchically related such that previously discovered basic knowledge should serve as an input into more detailed versions.

When constructing a model of a complex system, the purpose of the model is important. In physics we know that any model does not come even close to capturing all the details of the system. Therefore, we have become accustomed to the idea that ``the model should be as simple as possible but not simpler", but we want the model to describe some basic features or behavior of the real system, at least reasonably well. Thus in our model building we aim for tractability and clarity, by considering that `models are like maps' so that they are useful when they contain the details of interest and ignore others.{\footnote{A passage from Lewis Carroll's Sylvie and Bruno Concluded illustrates this point: ``What do you consider the largest map that would be really useful?" ``About six inches to the mile." ``Only six inches!" exclaimed Mein Herr. ``We very soon got six yards to the mile. Then we tried a hundred yards to the mile. And then came the grandest idea of all! We actually made a map of the country, on the scale of a mile to the mile!" ``Have you used it much?" I enquired. ``It has never been spread out, yet," said Mein Herr: ``The farmers objected: they said it would cover the whole country, and shut out the sunlight! So now we use the country itself, as its own map, and I assure you it does nearly as well."}} So the utility of simple models in describing the complexities of, for example, poorly understood ICT-based social systems is very high. Simple models may give deep insights in the same way that the simple Ising model provides useful understanding and quantitative correct predictions on critical phenomena of real magnetic systems.

The complexity of a system appears as emergent properties in its often complicated structure, in how it functions, and in how it responds to external influence of different kinds. These properties can best be studied empirically from the perspective of data analysis. While in natural science well-devised experiments produce the necessary data, in social science the rapidly increasing availability of large scale datasets or ``digital footprints" left by humans in various ICT-related systems and services has created the development of data- or reality-mining (which can also considered as statistical inference). This has become the main source of empirical studies. These should be accompanied by appropriate modeling. In this respect it is important to note that the term `modeling' has different meaning for physicists, statisticians and social scientists. While in physics modeling is mainly aimed at the understanding of the underlying mechanisms, in social science (and in statistics) a fitted curve is already considered to be a model. Here we use the term more in the physics sense.

Modeling social complex systems constitutes a major challenge because of the multiple scales and facets involved. To make progress we need to integrate various research approaches of different science disciplines. These include: social psychology experiments and surveys, data mining, network analysis, complex system and network theory, agent-based modeling and game theory, theory of phase transition and critical phenomena, intelligent and automated (ICT-based or radio-frequency) data-collection systems {\it etc}. These different approaches when integrated and fused yield a more comprehensive and complementary picture of the underlying social mechanism and social dynamics at different size-scales from individual, to dyad, to triad, to group, to community, and to whole society level. Moreover, it is our belief that the future of social experimenting lies in the combination of computational and experimental approaches, where computer simulations optimize the experimental setting and experiments are used to verify, falsify, or improve the underlying assumptions of the model.

The aim of simple models is to get better understanding of the so-called ``stylized facts" of the system, {\it i.e.,} to make simplified, abstracted, or typical observations - in other words capture some ``essence" of the real system. Of course simple models do not describe all the details of a system under consideration. Another possible advantage of simple models is that they may facilitate an analytic treatment and, thereby, give better insight to the plausible mechanism explaining the behavior of the system of interest. Simple models can be extended or made more complicated in a step-by-step way to capture more details of the system of interest. Moreover, simple models may be very useful in proving that statements made of a system are wrong, {\it i.e.,} they have an eminent role in the falsification process. On the other hand it may turn out that the model describes what everybody already knew, {\it i.e.,} some common wisdom. In this case the model, though simple, captures some of the salient features of the real system. This then serves as a starting point for more complicated models, with the hope of capturing even more features of the real system of interest correctly.

Concerning the predictive power of models, it is not necessarily the case that more complicated models do a better job. In fact it often turns out that simple models can do a very good job due to their clarity and tractability. Therefore prediction or forecasting capability is not always a good measure for the usefulness of models, but rather testable model implications are pluralistic\cite{pluralistic}. Furthermore a distinction between prediction and forecasting should be made: prediction should carry a weaker but more general meaning, {\it e.g.} by predicting {\em types} of behavior rather than quantitative forecasting. Also forecasting models, such as weather forecasting, are often based on known physical laws. In case of social systems one is more inclined to think the aim of computer simulations is to predict qualitatively the possible behavior of the system.

Being able to predict the behavior or forecast the dynamics of a system is followed by the possibility of it being managed. Once one is able to model and predict the behavior of the system, even qualitatively, it yields understanding of the system enabling the making of decisions, policies and further development of the system. This constitutes the ability to manage the system at some level, which can further be enhanced by improving the models step-by-step. If on the other hand one is able to go in more quantitative directions and forecast system outcomes, the model can then serve as a tool for developing and optimizing the system and its functions.\\

\noindent
\textbf{Micro-macro connection. Choice of variables.}

\vspace{0.08in}

There are instances in which a well established methodology exists to link the micro and macro descriptions allowing an appropriate choice of variables to describe a given phenomenon. This is not generally the case and a methodology is needed to avoid the temptation of ultrarealistic models in which irrelevant information is included: for example, sub-nuclear, atomic or molecular description of water is useless for wave motion in the sea; and particular car engine characteristics are not relevant in traffic modeling. Examples of these methodologies include the connection between the atomistic and hydrodynamic descriptions as successfully used in traffic modeling (the cars being the atoms) or the choice of order parameters based on symmetry principles as the appropriate variables for the study of continuous phase transitions. Another example is the methodology of the Renormalization Group in Critical Phenomena that provides a mathematically framework to identify relevant and irrelevant parameters by looking for an analytical description of the systems in the space of the scale transformations. This also leads to a well defined and operational meaningful notion of universality. Likewise, centre manifold theory allows one to identify the relevant variables and to derive their dynamical equations (amplitude equations), through a multiple time scales analysis, for the description of a system close to an instability point. Some of these methodologies appear in different contexts with different degree of mathematical formalization, but with the same basic contents, such as the derivation of amplitude equations or the \emph{slaving principle} of Synergetics (justified rigorously by normal hyperbolicity theory).

These examples show that success is possible, but finding a framework to solve this question in general complex systems remains a challenge. The rationale behind this question was spelled out by T. Schelling \cite{Schelling} in his \emph{Micromotives and Macrobehavior} book:\emph{There is a class of important propositions that are true for the aggregate and not in detail, and that are true independently of individual behavior}. Of course this does not refer to simple statistical properties of a large number of independent units, but to emerging phenomena that result from their non linear interactions.

The challenge remains to find a methodology or a classification of methods and protocols for the choice of variables describing complex behavior. Better choices than intuition or focusing on the variable of interest for the observer are needed. Many financial market models consider that all relevant information is contained in prices, and therefore there is no need to consider anything else. But is this the only relevant variable? In models of opinion dynamics the preferred variable of choice is the proportion of people with a given opinion, but it might well be that this is not the most relevant variable of the dynamical process which should  be extracted {\it a posteriori}, in the same way that a directly observed quantity in a physical instability is not the dominant amplitude variable for which the dynamics is well characterized.\\

\noindent
\textbf{Beyond the emergence of simple collective behavior}

\vspace{0.08in}

Simple collective behavior is an emergent property in the behavior of an aggregation of interacting units that cannot be understood from extrapolation of the properties of the units. This, for instance is the case of phase transitions in physical systems. There are well established theories and concepts like \emph{broken symmetry} \cite{Anderson} to understand these situations. Flock formation is another example of simple collective behavior \cite{vicsek_1995}. But already in his pioneering paper of 1972 Anderson identifies that \emph{the next stage could be hierarchy or specialization of function, or both} \cite{Anderson}. Indeed, there are emergent phenomena that, beyond not being reducible to individual properties, give rise to hierarchy, multilayered structures and functionalities - a prominent example being the emergence of organizations and institutions in social systems. We are still lacking a general theory or a satisfactory and sufficiently general conceptual framework to describe and understand these emergent properties.

\vspace{0.08in}

\noindent
\textbf{Beyond correlations: the search for cause-effect relations}

\vspace{0.08in}

Many studies of what are today considered as complex systems have traditionally relied on blind statistical analysis. The observation of these systems provides correlations of different types. Sometimes these correlations are considered to be some type of ``laws of nature" that should be reproduced by {\it ad-hoc} modeling.
To go beyond the knowledge provided by these correlations and to be able to establish cause-effect implications
is an urgent challenge.
This general question appeared long ago, for instance in the economics literature \cite{Granger}. Recent work in this direction is in the context of directed networks inference \cite{Kenett2010,Hempel2011}. Still we are far from a satisfactory solution to this question. On the one hand it requires the identification of mechanisms that are isolated and implemented in models to investigate their consequences. On the other hand it also requires new approaches to data gathering and analysis.

Common sense thinking and problem solving often adopts the concept of a single cause and a single effect. It also suggests that small changes in the cause imply small changes in the effect. It does not literally mean that there is a linear relationship between the cause and the effect, but it means that the system's behavior will not be surprising, and it is predictable, {\it i.e.,} changes in the parameters or in the structure of the system do not qualitatively alter its behavior, and the system is structurally stable.

Circular causality in essence is a sequence of cause and effect whereby the explanation of a pattern leads back to the first cause and either confirms or changes that first cause. The concept itself had a poor reputation in legitimate scientific circles, since it was somehow related to use vicious circles in reasoning. It was reintroduced into science in cybernetics \cite{wiener} emphasizing feedback. The concept of circular causality is reflected also in the theory of reflexivity, an approach promoted in economics by George Soros \cite{soros}.\\


\noindent
\textbf{Data}

\vspace{0.08in}

Empirical science is based on the analysis and modeling of data. The explosion-like development of ICT has resulted in an enormous increase in the data available for investigation. This is true for traditional ``hard sciences" but even more so for the social sciences. Many of our activities leave digital footprints that form huge data sets. Our phone and email communication, browsing the internet, using applications like Facebook, and commercial activities are all documented and can be used for scientific analysis to provide insight into phenomena and processes at the societal level. This approach has already made it possible to understand the relationship between the structure of the society and the intensity of relationships, the way pandemic diseases spread and what are the main dynamic laws of human communication behavior.  The availability of data makes it possible to study in detail some of the most intriguing aspects of complex systems, namely their hierarchical structure, and how it is related to the dynamics. What are the laws of ``microscopic" social interactions? How meso-level structures form and what is their role? What are the emergent cooperative phenomena at the societal level?
ICT has enabled a new approach called computational social science and this puts these questions within the scope of empirical investigations. The extension of empirical analysis to include massive ICT data, supported by large-scale multi-agent modeling
has provided
social science with immediately applicable tools able to handle issues of major concern. In fact, it gives the hope that mankind may be able to cope with many pressing issues.

The data deluge related to the ICT brings up several challenges. First, much of the data are not publicly available. Some of them, like mobile phone data are company property, while others such as much financial data are only available commercially. There have been efforts by scientists to create an openly accessible pool of data for research purposes \cite{Jerusalem,Science}, but perhaps the most severe problem of data driven social science is related to this point. In ``hard sciences" reproducibility of results is crucial, but without open access to data this is not possible in computational social science. While we should aim at broadest possible availability of data, the production of  well calibrated artificial data sets is one of the important tasks in this field.

Further challenges are related to the quality of data. Most often ICT related data are not collected for scientific use but, {\em e.g.}, for billing purpose as in the case of mobile phone data or for marketing like in the case of point collecting in supermarkets. In such cases metadata like age, gender, location etc. of the people can be assigned to the data in a rather noisy manner. This leads to impure data, with gaps and mistakes. Cleaning the data and constructing reference data or standards need new techniques and here massive interaction with social scientists will be necessary.

Of course, the handling of sensitive data raises ethical problems \cite{HelbingData}. At the moment there is only limited regulation in this respect. Two opposite opinions have been formulated: (i) no extra regulation is needed in addition to the general legal framework; and (ii) there is need for institutional solutions similar to those in genetics and traditional social science. A thorough study by the National Academy of Sciences, US \cite{NATUS} supports the latter view.

Even in everyday life the data deluge has changed our attitude to information. While previously {\em searching} for information was typical, today {\em selection} has become most important. This means in science that data mining and processing techniques have become crucial for the development of the field of complexity science. While the question of dimensional reduction of the data to arrive at useful information remains a central problem (as is usual in empirical science), the next foreseeable frontier is a complementary approach that implies a shift from data-driven modeling to question-driven data-gathering {\it i.e.,} the goal is producing or gathering data to answer a specific question. This is in the spirit of classical experiments designed to obtain data to test theoretical predictions. True validation of models, as opposed to models fitting raw data, requires comparison of model implications (quantitative or qualitative) with data obtained under the conditions and assumptions of the model. Such experiments are naturally designed in some virtual environments, like internet games or in electronic social networks, and they are part of the new undertaking in social experimenting \cite{HelbingSimulation}. Of course, ethical issues have to be handled with appropriate care. We refer to Section~\ref{platforms} for a more detailed discussion of the platforms for social computing and web-gaming.

\vspace{0.08in}

\noindent
\textbf{Ensemble modeling and Data assimilation}

\vspace{0.08in}

There are fields such as weather prediction or climate research in which the fundamental microscopic laws are well known and established but given the huge range of relevant scales, meso and macro models are used that implement in different ways large-scale effective interactions and parametrizations of the same basic phenomena. In these fields a common practice involves probabilistic forecasts based on the combined results of different models or different specification of a given model. This methodology is used in a variety of fields \cite{Araujo}. An important methodological question is to which extent this pluralistic modeling \cite{pluralistic} or combination of forecasts \cite{Bates} is conceptually justified beyond purely statistical considerations, in other fields, such as social phenomena modeling.
Pluralistic modeling should, however, not mean that different models originating form different basic concepts can be simultaneously considered.
A related methodological question is the possible general use of the \emph{data assimilation} procedures of atmospheric modeling \cite{Kalnay}. The basic idea is combining forecasting and observation for initial conditions in dynamical modeling: Forecasting at a given time has to be combined with large scale observations at that time. Another methodology recently developed in climate (see e.g, \cite{Gozolchiani}) is based on analyzing the hidden information in the dynamics of weather similarity (network links measured by cross-correlations) between pairs of locations. This yields an evolving network characterization of the climate that was found useful for example in better understanding of the El-Nino phenomena.

\vspace{0.08in}

\noindent
\textbf{From Data to Dynamical Models}

\vspace{0.08in}

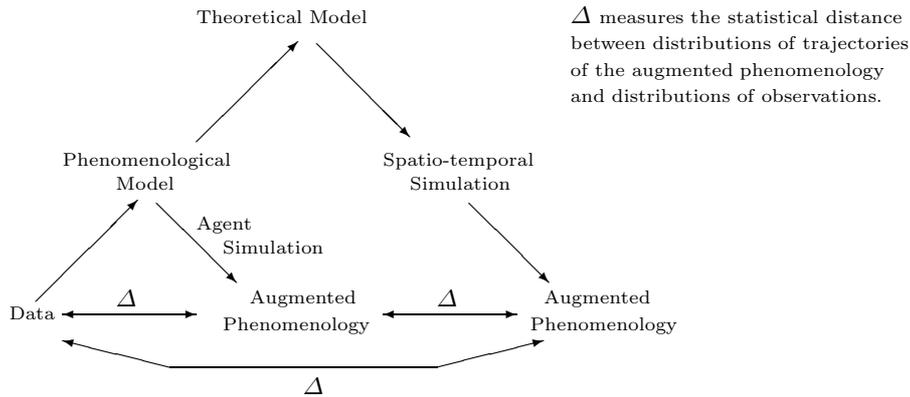
\begin{figure}[t]
\begin{picture}(300,190)(50,-35)

\put(290,130) {$\Delta$ {\scriptsize{measures the statistical distance}}}
\put(290,120) {\scriptsize{between distributions of trajectories}}
\put(290,110) {\scriptsize{of the augmented phenomenology}}
\put(290,100) {\scriptsize{and distributions of observations.}}

\put(150, 130) {\scriptsize{Theoretical Model}}

\put(100, 76) {\scriptsize{Phenomenological}}
\put(120, 67) {\scriptsize{Model}}

\put(220, 76) {\scriptsize{Spatio-temporal}}
\put(230, 67) {\scriptsize{Simulation}}

\put(150, 85) {\vector(1,1){38}}
\put(195, 122) {\vector(1,-1){35}}

\put(252, 62) {\vector(1,-1){30}}

\put( 90, 25) {\vector(1,1){38}}
\put(135, 62) {\vector(1,-1){30}}
\put(150, 52) {\scriptsize{Agent}}
\put(160, 43) {\scriptsize{Simulation}}

\put( 80, 18) {\scriptsize{Data}}
\put(100, 20) {\vector( 1,0){50}}
\put(150, 20) {\vector(-1,0){50}}
\put(170, 24) {\scriptsize{Augmented}}
\put(160, 14) {\scriptsize{Phenomenology}}
\put(220, 20) {\vector( 1,0){50}}
\put(270, 20) {\vector(-1,0){50}}
\put(280, 24) {\scriptsize{Augmented}}
\put(275, 14) {\scriptsize{Phenomenology}}

\put(140, 0) {\vector(-4, 1){40}}
\put(240, 0) {\vector( 4, 1){40}}

\put(140, 0) {\line( 1,0){ 100}}

\put(190,-10) {$\Delta$}
\put(120, 23) {$\Delta$}
\put(240, 23) {$\Delta$}

\end{picture}
\caption{The complex systems methodology for reconstructing models and theory from data}\label{figure-1}
\end{figure}

Figure \ref{figure-1} illustrates
the scientific perspective of complex systems methodology.  It begins with data from which scientists reconstruct phenomenological models. For example, Kepler constructed a phenomenological model in which the planets sweep out equal areas in equal times which Newton formulated as a theory of planetary motion able to reproduce this phenomenology. In the case of the motion of two bodies, Newton's Laws produce equations that can be solved explicitly making it possible, for example, to predict precisely where a cannon ball will land. In the three-body case the equations cannot be integrated and the system is chaotic. Nonetheless the spatio-temporal behaviour of the system can be simulated by iterated computation providing an augmented phenomenology (Figure \ref{figure-1}, bottom right). The objective in this modeling is to produce an augmented phenomenology whose statistical difference from observation, $\Delta$, is as small as possible (theoretically zero for a perfect model). In most cases simulations can at best sample the space of all system trajectories around given initial conditions with an error, $\Delta$, which measures the difference between the statistical distributions of the simulated trajectory and the statistical distributions of the data. Since each iterated calculation in a simulation of a system sensitive to initial conditions creates error, $\Delta$ increases with time.

As a social systems example consider the people evacuating a building in an emergency. Helbing \cite{helbing2000} observed the motion of people in crowds and created a phenomenological model of the ways people move with respect to each other. Using this phenomenological model Helbing used agent-based computer simulation to create an augmented phenomenology for this system (Figure \ref{figure-1}, bottom-centre). Helbing went on to formulate theoretical models of pedestrian flows (centre top) permitting spatio-temporal simulations (mid right) to create another augmented phenomenology (Figure \ref{figure-1}, bottom-right).
In this case both the agent simulation and the theoretical model gave an augmented phenomenology with small error. In fact Helbing went on to use this new science to assist the authorities to redesign Mecca for the Hajj pilgrimage which was subject to fatal accidents with large numbers of people being trampled as the dynamics of the crowd changed.
The redesign was very successful and many lives have been saved \cite{helbing2007}.
This is one of the major success stories of complex systems science.

\section{Interconnected multiple scales networks}
\label{Diaz-Guilera}

In 1998 Watts and Strogatz \cite{ws98} presented the first evidence of what they called {\it complex networks}. Watts and Strogatz realized that patterns of connections in the neural network of a worm, in the power-grid of the United States, and in the network of co-appearances of actors in movies, showed common features that could not be explained by simple mathematical models (the two extremes of total order or total randomness being the most simple). The presence of short-cuts made these networks extremely small while at the same time kept track of their ordered origin through a larger than expected clustering coefficient (related to the existence of triangles in the network). On the other hand, as shown slightly later by Barabasi and Albert \cite{ba99}, the distribution in the number of connections coming out from given node ({\it e.g.} computers in the Internet, pages in the world-wide-web, or number of papers authored by scientists) show skewed distributions, which means that there exist some actors in the networks that are highly connected, which were called {\it hubs}. These two works were the seed of a new field of research.

Nowadays, it is understood that most complex systems show emergent dynamical properties which are inherently related to the topology of the underlying network of connections among the constituent parts of the system. During these fourteen years we have witnessed how these ideas have been applied to a myriad of problems ranging from the cell scale of biochemical networks to the scale of world population communicating through the information and communication technologies \cite{blmch06,bbv08}.

Up to now we have considered mostly that a given network description is good for a given problem. If we plan to understand how proteins interact then we look at the protein-protein interaction networks, and if we want to prevent damages in the energy distribution we analyze the features of the power-grid networks across the globe.

But the question that arises now is to which extent these networks are really independent? Is the distribution of electricity across the power-grid independent of the transportation of other energy resources such as gas \cite{cbbgja09}? Is it independent of the trade relation between countries as illustrated by the recent diplomatic issues between former soviet republics? Is it independent of the communication network that connects power stations and distribute energy according to generation and load? The answer is clearly not - these networks are not independent. The communication network among power stations depends on the stability of the power-grid and the same for the other examples \cite{bppsh10}. Buldyrev et al \cite{bppsh10} recently introduced a mathematical framework, based on percolation theory, to study the robustness of a network formed by interdependent networks.

But this interrelation not only affects technological or transportation networks, it also affects social networks. To how many social networks do we belong? From the traditional classification of social networks we can identify friendship networks, kinship networks, professional networks, and in a more modern framework we belong to different types of online networks. Networks can relate customers among themselves: people who purchased the same items, people that like and comment on each other's pictures, and people cooperating in online games together. How are all these networks intertwined and how do they affect each other? Online social networking is changing our way of living, sharing, or even feeling. Perhaps one of the main drawbacks of Facebook is that the definition of a ``friend" just means someone you are connected to. Whenever you, as Facebook user, receive a message saying: ``... wants to be your friend", it can be your mother, your student, your thesis advisor, an ex-girl (or boy) friend, or your online war allies. On the contrary, a new proposal by Google (Google+) tries to change this feature with its ``circles", in which one can choose the type of relationship with the other users and hence separate the different networks we belong to. Those are clearly two different perspectives on something that for people is simple ``who is connected to who", but for scientific reasons it is extremely important to distinguish these types of connections that make up our social world.

Even social and technological networks are strongly interconnected, {\em e.g.} how much our connected social world depends on electricity resources. Among the most studied networks over the years we find the Internet and the World-Wide-Web. The first is identified with the set of hard wired connections (and now includes wireless protocols as well) between Internet providers and users that forms the basic infrastructure for all current communication technologies. The second corresponds to the set of pages which contains the information. But nowadays, the flow of information is not only provided in one direction from the web servers to the final users, but strengthens relations among users or among communities at different scales through different ways of communication: emails, chats, messaging, blogging, sharing, and so on. But, closing the circle, new social interactions are still on top of technological networks that are vulnerable and can be monitored and even censored.  And one of the most clear examples took place during summer 2011, since messages fueling the London riots were mainly broadcast from
%
smartphones using a private network which encrypts messages.

Additionally, these networks form aggregates of users that can be analyzed at different scales, because the questions one poses and hence the answers one gets are quite different. Studies have been performed on online users that are in the same class-room or at college level \cite{herrmann}. Other studies focus on the spreading of political ideas through Facebook or Twitter (in the Arabian countries, in Finland or later in Spain and the 15M movement of ``the indignados" \cite{15M} (http://15m.bifi.es/). Finally, other empirical analyses have focused on the widespread use of messaging services as Microsoft Messenger \cite{messenger}.

This importance of different scales is also observed in the propagation of diseases. Spreaders (humans) can travel long distances following different transportation modes: long and fast jumps made by flights through the airport networks; medium range trips are mostly by car or bus then related mainly to road networks; while short and frequent trips correspond mainly to commutation networks of public transportation. In this case transportation networks at different scales interact. This is how real infections spread, but a curious example about the relation between networks of different origins is found in the records from the web searches during the H1N1 virus propagation. These searches, made by people with some of the symptoms are geolocalized and combined give hints on how the disease is being propagated.

In the near future  researchers on complex social networks will focus on networks at different scales, with intertwined different meanings. Probably the picture will not be that simple. Nowadays we know that networks are directed, weighted, adaptive, space embedded, interdependent and so on, and furthermore they are also dynamic. This includes  the dynamics taking place over the links \cite{bbv08,adkmz08}, and also the inherent dynamical nature of the connections themselves, with coevolution processes of the dynamics {\em on} the network and the dynamics {\em of} the network \cite{coev1,coev2}. This will introduce further ingredients that the new science of complex networks will face in the future \cite{review_temporal_networks}.

For all these reasons, complex networks theory has to develop new tools, new measures, and new models that account for all these new ingredients, namely the interrelations between networks of different origins, networks interacting at different scales, cross correlation between dynamics on networks and the dynamics of network topology, including dependence on patterns of connectivity. Collaboration with social scientists and ICT-researchers will be crucial in developing a new framework for the final understanding of new social features and behaviors, and to construct new socially inspired communication resilient technologies. Also, for more details on current challenges on complex networks research see the paper on networks by Havlin {\em et al} in this volume \cite{Havlin2012}.

\section{Information aggregation and processing. Social learning}

At different social levels, from the family, to international coalitions, to the global human society, we need to take collective decisions that shape our future. Taking a good decision ultimately depends on our ability to aggregate information that is widely dispersed. This implies making choices on what information we pay attention to and how we discriminate between what we consider relevant or accurate information and what we consider background noise. These processes depend on very different issues, some of them strongly technological such as information propagation and information availability, and others strongly social such as trust. Needless to say, the problem of information aggregation takes a completely new perspective in the light of the societal changes associated with the new Information and Communication Technologies, which imply different, faster and broader ways of communication as described in the next section. Today the flow of information occurs at multiple temporal and spatial scales. The compelling task has changed from accessing information to selecting information avoiding misinformation and disinformation. How do we do this selection and aggregation, and what are the consequences?

From the above perspective, social learning can be defined as the ability of a population to aggregate information \cite{Gonzalez-Avella2011}. This process drives phenomena like opinion formation or political changes (either smooth or deep changes). Individual learning follows, in a traditional setting, from global-local competition: the competition between global information received through media, advertisements, {\it etc.} and information learnt and adapted from one's social circle. This competition has today different characteristics due to the new extent and meaning of the social circle: the ease of interacting with any other individual redefines the concept of social circle. An important change of attitude also exists with respect to news information: many people now search actively for news in different ways on the internet instead of passively watching TV news or habitually reading the same newspaper. The challenge is to understand that processes of information aggregation in a society are not necessarily driven by information hubs that process information and broadcast it, but are strongly coupled in a society that aggregates information in different collective processes driven by pair-wise, group and multi-institutional interactions. A clear example in this context is the change from the Encyclopedia Britannica to the Wikipedia.

A basic question is the possible shift from a society that exploits social learning by aggregated information and feedback processes to improve this learning, rather than relying on the traditional specialists or experts. An important challenge in this context is how to avoid or how to identify information cascades or rumor-spreading that amplify errors or misconceptions to a globally accepted false truth. There is the \emph{the wisdom of crowds}, but crowds are not always wise. Also, there is the question of different processes for aggregation of information about facts and about interpretation of the facts, that is the aggregation of meaning. Collectively these questions are behind two of ten top social-science questions listed in \cite{GilesNature}: \emph{How can we aggregate information possessed by individuals to make the best decisions?}, and \emph{How can humanity increase its collective wisdom?}

Information aggregation and processing is not specific to human societies and is a common process in many other natural complex systems, as for example in bird flocking or fish schooling. In this context a recent experiment \cite{WardPNAS} has demonstrated genuine \emph{wisdom of the crowd} \cite{Conradt} showing that larger shoals of fish make more accurate and faster decisions (avoiding a predator) than smaller shoals. The design of the experiment is also an example of data gathering to test modeling predictions or to answer a given question, beyond the use of raw available data. What we learn about information processing by decentralized information communication in natural systems is also a guide to designing new and alternative information processing systems, searching for the emergence of intelligent behavior from simple interaction rules. Research in Swarm Intelligence is an example in the direction of implementing self-organized coordination of many individuals by decentralized information communication.

Finally, knowledge aggregated from complex self-organizing human or natural systems opens up the challenge of the implementation of unconventional computational principles based on complex dynamical systems \cite{Crutchfield2010}, such as \emph{reservoir computing} \cite{Jaeger,Bunomano} or information processing based on complex systems dynamics  \cite{Appeltant}.

\section{Socio-Technical Systems}

The term ``socio-technical systems" refers to the interaction between technologies and human social behavior. Psychologists initially recognized this interaction in the early 50's. In a pioneering work E.L. Trist and K.W. Bamforth \cite{Trist1951} studied the social consequences of the adoption of a new production technology in coal mining leading a productivity fall. In this general context one can ask what happens to a society when new forms of communication appear. This question, which is fundamentally important and has far-reaching implications, is what the Information and Communication Technologies (ICT) has brought about over the last decade. ICT has radically and unforeseeably changed society as a whole. This is true not only in highly industrialized countries as shown for example by the large impact and penetration of mobile phone networks in developing countries in Africa. At first sight, these changes can be attributed to the actions of individuals and the availability of new channels of communication that transform basic social processes: (i) face-to-face encounters have become less critical than in the past, (ii) the dynamics of building and strengthening relationships have evolved by taking advantage of ICT, and (iii) new ICT-mediated groups and communities have emerged, by overcoming typical limitations such as distance or lack of a common platform. In addition, entirely new ways of collective human behaviour have appeared, such as those collaborative and sometimes conflicting actions exemplified by Wikipedia.

However, this description is critically incomplete because it fails to recognize that individuals, society, and ICT are deeply intertwined in a dynamic feedback process, where individuals adopt new communication channels to form and join groups that change in identity and size, thereby restructuring the whole of society. Simultaneously, ICT providers develop new channels of communication, some of which fail while others become enormously popular. Indeed, unpredictability is a characteristic feature of these developments. Popular channels such as WWW and SMS were not originally designed for the purposes they serve today. Entirely new platforms for ICT-mediated social interactions, for example Facebook, have emerged ``out of the blue". They have gained mass popularity in a very short time and transformed the social behaviour of individuals in a number of unexpected ways. An example is the role of Twitter in mass movements such as Arab Awakening of 2011 or the Spanish 15M movement \cite{15M}. In our view, the fundamental challenge for future social ICT is to overcome the acute lack of understanding of the driving forces and mechanisms of this complex system of interactions between individuals, society, and ICT.

This deficiency requires developing systematic means of exploring, understanding, modelling and possibly even controlling systems where ICT is entangled with social structures. In particular, there is need to focus on the behavioural patterns, dynamics and driving mechanisms of social structures whose interactions are ICT-mediated, from the level of individuals, dyads, and triads  to the level of groups, communities, and large-scale social systems \cite{Grabowicz2012}. The research approach necessarily has to be based on combined expertise in complex systems, computational analysis and modelling, and social sciences. In contrast with studies that start from extremely simplified assumptions concerning social dynamics and concentrate on finding structural features of social systems, it should be emphasized that ICT networks are dynamic systems of interacting humans and groups, and should thus fully utilize the theories and methods of the social sciences.

New ICT also puts public goods problems in a new perspective. An example is water management or waste management. How do we take decisions on our individual behavior in these issues in a globally interdependent society when we receive full information on daily situation and global consequences? Likewise, smart grids are now designed so that centralized decisions for better energy management are taken in the context of data collected from a large number of distributed sensors. The challenge is to introduce into the design the adaptive behavior of the users and to explore self-organized grids in which properly aggregated data from the sensors is made available online also to the users \cite{Johnson2009}.

\section{Data gathering, Participatory Sensing and Social Computing}

\subsection{Citizen Science}
\label{citizen_science}

The issue of sustainability is now at the top of the political and societal agenda and is considered to be of extreme importance and urgency \cite{Gore_2007,Randers_2004}.  There is now overwhelming evidence that the current organisation of our economies and societies is seriously damaging biological ecosystems and human living conditions in the very short term, with potentially catastrophic effects in the long term.  In a recent statement from the head of the European Environmental Agency, there is a realisation that only through bottom-up actions we can deal with today's challenges: ``The key to protecting and enhancing our environment is in the hands of the many, not the few.... That means empowering citizens to engage actively in improving their own environment, using new observation techniques...'' \cite{EEA_2009}.

The enforcement of novel policies may be triggered by a grassroot approach, with a key contribution from information and communication technologies (ICT). Nowadays low-cost sensing technologies allow the citizens to directly assess the state of the environment; social networking tools allow effective data and opinion collection and real-time information spreading processes. In addition, theoretical and modeling tools developed by physicists, computer scientists and sociologists have reached the maturity to analyse, interpret and visualize complex data sets.  A techno-social system, acts like a lens that captures information from the environment: one has to explore the peculiarities of having human agents as sensing nodes, the role of noise sources at different scales, the effect of opinion bias, information spreading in the community supporting the techno-social system, network effects, and so forth.

Devices employed in the connection to communication networks have converged in size and technological standards. Cell phones have integrated many functions traditionally accomplished by personal computers.  This progress while being useful, yields also new kind of risks and challenges such as epidemics of viruses and malfunctions\cite{barabasi_science_2009}.  In turn, computer manufacturers have privileged products designed for an easy mobile usage, such as new generation tablets. Moreover, cell phones and PCs incorporate sensors of increasing accuracy: GPS sensors, cameras, microphones, accelerometers, thermometers are already standard equipment in many devices.  Networks have also accompanied this process, by expanding the availability of an Internet connection throughout daily life. Open-hardware platforms, such as the well-known programmable microcontroller based Arduino, will also facilitate the task of taking an input signal from the environment, process it, and deliver it through the Internet at a low cost.

The large number of sensors deployed is already turning urban areas into ``smart cities'', that is, intelligent and complex organisms able to process the sensors signals, visualise them and possibly trigger the automatic execution of appropriate actions\footnote{\url{http://www.urbanlabs.net/index.php/UrbanLabs$\_$OS$\_$(English)}} (see Michael Batty et al. contribution in this volume). The mobile, powerful, and permanently connected equipment described above makes any citizen a potential source of sensor data about her/his environment, with little or no scientific skill required.  Participatory sensing experiments involve communities of such individuals in the monitoring of a particular issue, e.g.  the quality of a metropolitan environment \cite{Paulos_2007} or the redevelopment of urban areas. This is not entirely new, since numerous ``citizen science'' initiatives have been already launched in areas ranging from ornithology to astronomy, with or without the help of sensors. A recent trend is represented by the integration of crowdmapping and participatory sensing through the web and several important initiatives have been carried out, {\em e.g.} to monitor the spreading of the Influenza A virus\footnote{\url{http://www.influweb.it/}} or social mobilization\cite{red_ballon}. It is important to remark how this data gathering activity is very relevant for the so-called data-driven simulations, {\em i.e.} simulations of complex systems whose predictability accuracy crucially depends on the interplay between the goodness of the modeling scheme and the possibility to monitor several observables to recalibrate in real-time the evolution of the system under investigation. In addition, online platform, such as \url{www.pachube.com}, have shown in practice how the data collection activity and its visual representation reinforce themselves.  The access to both personal and community data, collected by users, processed with suitable analysis tools, and re-presented in an appropriate format by usable communication interfaces, has the potential of triggering a bottom-up improvement of collective social strategies as well as stimulating fundamental shifts in public opinion with subsequent changes in individual behaviour and pressure on policy makers\cite{EveryAware}. Particular events, such as the nuclear crisis following the 2011 earthquake in Japan, have demonstrated that involving citizens in the environmental monitoring activity is an effective method to build accurate risk maps.  The participation of users in the monitoring affects both the resolution and the quality of the data collected. While traditional sensing generally involves a small number of highly controlled observation points, distributed sensing relies on the possibility of gathering large amounts of data from many uncontrolled sources, which cannot ensure high data quality standards; however, by means of statistical methods together with the possibility of storing and post-processing large datasets, this quality gap with respect to traditional sensing can be overcome.  Therefore, the analysis tools should be able to detect and filter out deviations due to sensors misuse or to biases introduced by the users themselves.

\subsection{Platforms for ICT-based experiments}
\label{platforms}

In the last few years the Web has been acquiring the status of a platform for social computing, able to coordinate and exploit the cognitive abilities of the users for a given task. One striking example is given by a series of web games \cite{vonahn_2006}, where pairs of players are required to coordinate the assignment of shared labels to pictures \cite{vonahn_2004}. As a side effect these games provide a categorization of the images content, an extraordinary diffcult task for artificial vision systems.  More generally, the idea that the individual, selfish activity of users on the web can possess very useful side effects, is far more general than the example cited.  The techniques to profit from such an unprecedented opportunity are, however, far from trivial. Specific technical and theoretical tools need to be developed in order to take advantage of such a huge quantity of data and to extract from this noisy source solid and usable information. Such tools should explicitly consider how users interact on the web, how they manage the continuous flow of data they receive, and, ultimately, what are the basic mechanisms involved in their brain activity. In this sense, it is likely that the new ICT-mediated social platforms, could rapidly become a very interesting laboratory for social sciences. In particular we expect the web to have a strong impact on the studies of opinion formation, political and cultural trends, globalization patterns, consumers behavior, marketing strategies.  A very original example is represented by Amazon's Mechanical Turk (MT) (\url{https://www.mturk.com/mturk/welcome}), a crowdsourcing web service that coordinates the supply and the demand of tasks that require human intelligence to complete. It is an online labor market in which users perform tasks, also known as Human Intelligence Tasks, proposed by "employers" and are paid for this. Salaries range from cents for very simple tasks to a dollar or more for more complex ones. Examples of tasks range from categorization of images, the transcription of audio recordings to test websites or games. MT is perhaps one of the clearest examples of the so called crowdsourcing and thousands of projects, each fragmented into small units of Work, are performed every day by thousands of users. MT has opened the door for exploration of processes that outsource computation to humans. These human computation processes hold tremendous potential to solve a variety of problems in novel and interesting ways.  Thanks to the possibility of recruiting thousands of subjects in a short time, MT represents a potentially revolutionary source for conducting experiments in social science \cite{Chilton_2009,Paolacci_2010,Huberman_2011}. It could become a tool for rapid development of pilot studies for the experimental application of new ideas. As a starting point for this new idea of experiments, the blog \url{http://experimentalturk.wordpress.com/} already presents a review of the results of a series of classic game theoretical experiments carried out on MT \cite{Suri_2010}.  Despite its versatility MT has not been conceived as a platform for experiments.  This is the reason why it is important to develop a versatile platform to implement social games. Here the word game is intended as an interaction protocol among a few players implementing a specific task and it is used as a synonym of experiment. The development of such a platform has to satisfy a certain number of requirements among which high modularity and flexibility, synchronous ({\it i.e.,} real time) and asynchronous interaction modes, robusteness with respect to heavy loads to process and store a continuous data flow. The advantage of this kind of experiments is that every useful piece of information and detail of the evolution will be fully available and leveraged for benchmarking as well as for the modelling activity. Moreover the effects of social interactions can be observed with a larger statistical basis and in a more controlled environment.  It should be stressed that these ICT-based experiments are truly general purpose since through them one can investigate complex phenomena in a wide range of disciplines including (but not limited to) social sciences, economics, psychology and linguistics. In the framework of European project {\em EveryAware} \footnote{\url{www.everyaware.eu}} a first prototype of such a platform is being realized, dubbed Experimental Tribe (\url{www.xtribe.eu}) (ET). ET is intended as a general purpose platform that allows the realization of a very large set of possible games.  It has a modular structure through which most of the complexity of running an experiment is hidden in a complex Main Server and the experimentalist is left with the only duty of devising the experiment as well as a suitable interface for it. In this way most of the coding diffculties related to the realization of a dynamic web applications are already taken care by the ET Server and the realization of an experiment should be as easy as constructing a webpage with one of the many online services for it. The benefit is twofold: on the one hand, it allows virtually any researcher to realize his own experiment with minimal effort, paving the way of the use of the web as a standard ``laboratory'' to perform experiments. On the other hand, it can be a strong ``basin of attraction'' for people willing to participate to experiments, making in this way recruitment much more easier than for single-experiment platforms.

\section{Systemic risk, extreme events and predictability}

Extreme events both in nature and society, such as earthquakes, landslides, wildfires,
stock market crashes\index{stock market crash}, the destruction of very tall tower buildings, engineering failures, outbreaks of epidemics {\it etc.}  may appear to be surprising phenomena whose occurrence does not follow any rules. Of course, such kinds of extreme events
are rare,  but they influence our everyday lives dramatically. Can we understand, assess, predict and control these events?

Complex systems theory offers a new perspective to understand the mechanism of the emerging patterns. As a consequence of natural and social crises, the occurrence of rare large extreme events are now the focus of extensive mathematical analysis \cite{Sornette2003,Albeverio2006}.

\subsection{Widening the Limits to Prediction of Extreme Events}

It is common to hear questions such as ``what is the probability of having a big earthquake in Iceland within a year?'' or ``how large might a possible stock market crash be tomorrow?''. The study of earthquake eruptions, the onset of epileptic seizures, and stock market crashes traditionally are investigated by very different disciplines which differ very much in their scientific culture.  The complex system approach emphasizes the similarities and offers some common methods to predict the behavior of these systems, and/or understand the inherent limits of their predictability \cite{Erdi}.

Standard statistical procedures neglect data points deviating greatly from others, the so-called called {\em outliers}. Extreme value analysis uses statistical methods to analyze rarely occurring events. Typically, extreme events occur in the tails of probability distributions as a function of the ``size'' of the events (such as energy, duration {\it etc}).  Emil Gumbel (1891-1966) a famous pacifist, contributed significantly to the establishment of statistical methods to describe extreme deviations from an ``average" behavior. As he wrote: ``It seems that the rivers know the theory. It only remains to convince the engineers of the validity of this analysis.''

Extreme value analysis, a branch of mathematical statistics, estimates the probability of extreme floods, large insurance losses, market risk, freak waves, tsunamis, {\it etc}. While the Gumbel distribution shows a light-tail (exponential decay), other classes of ``extreme value distributions'' behave differently.  Distributions of earthquakes and avalanches have extreme value statistics described by power-law tails. These imply that extreme events occur much more frequently than expected.  For example, the crash of the stock market on Black Monday was a 35$\sigma$ event, where $\sigma$ is the standard deviation of the Dow Jones Index on a logarithmic scale. Knowledge of the size distribution of floods, storms, earthquakes is highly important for the insurance business and for the risk assessment of financial derivatives.

 In the context of power laws, Sornette has suggested the possibility of ``transient organization into extreme events that are statistically and mechanistically different to  from the rest of their smaller siblings'' \cite{sornier}. He calls these {\em dragon kings} where ``Often, dragon kings are associated with the occurrence of a phase transitions, bifurcation, catastrophe, tipping point, whose emergence organization produced useful precursors''.

The theory of complex systems suggests that extreme events may be predicted by detecting their precursors, and that there are methodological similarities for analyzing and modeling different ``critical events'' occurring in physical, biological and social phenomena.  There are initial promising results and many open problems.

\subsection{Dynamical models of extreme events}

To be able to control and manage extreme events we should understand the generating mechanisms (and generative models) of the phenomena.  One possibility is to say that big earthquakes are nothing else but small earthquakes that do not stop. The consequence is that these critical events would inherently be unpredictable, since they don't have any precursors. This approach is called \emph{self-organized criticality} (SOC) and was championed by Per Bak \cite{bak96}.  Self-organized criticality suggests that the same effect may lead to small, but also to very large avalanches, so the outcome is not really predictable. A famous toy model is the sand-pile model \cite{baketal87}.

According to Sornette's arguments \cite{Sornette2003} catastrophic events, or at least a class of them, result from accumulating amplifying cascades. Based on the hypothesis of this theory of \emph{intermittent criticality}, many stock market crashes are generated by a slow building up of ``subterranean forces'', and their precursors may be detected. Were this hypothesis true, the predictability of these events may be possible.

Uncompensated positive feedback can be a mechanism for crashes.  Positive feedback seems to be a general mechanism \cite{Sornette2002} behind the eruption of earthquakes, stock market crashes, hyperinflation, and epileptic seizures. The lack of the stabilizing effects of negative feedback mechanisms may lead to catastrophic consequences. If there are no mechanisms to compensate for the effects of higher-than-linear positive feedback, the processes lead to finite-time singularities.

In an economy there are many feedbacks that
drives it towards equilibrium between demand and supply.
There are many positive feedbacks, such as those due to our susceptibility to imitate each other's behaviour, and these can lead to explosive growth in prices, followed by the inevitable bursting of the bubble.  Equilibrium theory works well when negative feedback effects have stabilizing effects to positive feedback changes. While in normal situations the activities of buyers and sellers neutralize each other, in critical situations there is a cooperative effect due to the imitative behavior of everybody wanting to buy since everybody else has already bought, and the positive feedback is higher-than-linear. Such super-exponential increases, due to irrational expectations, cannot continue for ever and the increase is unsustainable. Consequently, it should be followed by a compensatory process, {\it i.e.,} a stock market crash.

\section{Control and management of complex global systems}

A pervasive challenge for complex systems science is to control or manage complex systems.  Here is a list of topical examples:
\begin{tabbing}
xxxxx \= xxxxxxxxxxxxxxxxxxxxxxxxxxxx \= xxxxxxxxxxxxxxxxxxxxxxxxx \= \kill \\
\>Financial System
\>Social Unrest
\> Economy\\
\> Health Service
\> Famine Relief
\> Epidemics\\
\> Electricity Pricing Schemes
\> Demographics
\> Climate\\
\end{tabbing}

The words ``control" and ``management" may be used interchangeably, but many social scientists dislike the word ``control" which carries overtones of authoritarianism and prefer the word ``management" which conveys a more benevolent approach.

There is an important distinction to make, however, between two forms of control.  In the strong form of control, the objective is to make the trajectory of the system follow some desired track or reach some target \cite{Sontag}.  In the weak form of control, the objective is to make the probability distribution for the trajectories of the system follow some desired track or reach some target (in stochastic control, some integral with respect to the probability distribution is usually optimised).

This distinction is crucial.  As soon as the dynamics of a deterministic system shows sensitive dependence on initial conditions (``chaos"), control of a trajectory is likely to require feedback with higher gain than the maximal Lyapunov exponent (though examples can be made where arbitrarily small carefully chosen gain matrices suffice), which may involve unrealistically high observational power and actuator response.  Similarly, control of a stochastic jump system requires control response time to be shorter than typical waiting times. In the case of diffusive systems, a similar criterion is required but depends on the accuracy with which one wishes to track.

In contrast, chaos or stochasticity are good for making the probability distribution for the trajectories of a system relax rapidly to one that is unique.  Thus this probability distribution may be controlled by much slower observation/actuator feedback than the individual trajectories.  Also much less detailed observations and controls may suffice, thus avoiding Orwell's ``big brother" nightmare and making them more acceptable to our ``free" society.

This section concentrates on the control of probability distributions for trajectories.  They have been christened ``space-time phases" \cite{Mac08}.  In the next few paragraphs we propose a role for substantial new mathematical developments.  For some background, see \cite{Mac11,DM11}.

The simplest context in which to begin is probabilistic cellular automata.  These consist of a network of units whose states update in parallel in discrete time according to probability distributions that depend on the current state of the whole network but, conditional on the current state, the distributions for different units are independent.  In contrast to much of the literature (e.g.~\cite{Liggett}), there is no assumption here that the network is a regular lattice, that the units are identical, or that the dynamics are autonomous.

Under suitable conditions, the operator representing the evolution of probability distributions for the state of the network is an eventual contraction in a suitable metric, and this leads to exponential convergence to a unique probability distribution for the trajectories.  The resulting space-time phase depends smoothly on parameters of the model, thus its dependence on feedback control laws can be studied.  Given design objectives, one could then seek feasible control laws to bring the statistical behaviour of the trajectories close to the objectives.

Although the above holds for all indecomposable systems with finitely many units, a more appropriate approach for large systems with some strong interdependence of their units is to consider them as part of an infinite system (just as in equilibrium statistical mechanics).  Then the possibility of non-unique space-time phase emerges.  As parameters are varied the system may jump from one space-time phase to another that is far away.  This is a reflection of the popular notion of ``tipping point".  Even without parameter variation, the original finite system may best be described as making random transitions between two or more such phases.  The ways the set of phases can depend on parameters is a fairly wide open question: some semi-continuity results hold, but there is a great need for an analogue of the bifurcation theory for simple attractors of deterministic dynamical systems, so that we could understand what are the typical qualitative changes in the set of phases.  Going further, could controls be designed to collapse the set of phases into a desired unique one?  This connects with another branch of deterministic dynamical systems theory called ``ergodic optimisation" in which the aim is to stabilise an invariant probability distribution on an attractor differing from that naturally chosen.

Once the theory for probabilistic cellular automata is well developed, it will be natural to seek to extend it to more realistic classes of system, for example continuous-time stochastic jump processes, systems of mobile units where the strength of interaction depends on distance in physical space, and deterministic systems with sufficiently chaotic dynamics.

Let us turn now to the examples.

Redesign of financial regulation is urgent.  The current space-time phase has bubbles and crashes all the time, even if the recent banking crisis has been the worst ever and national debt crises may overtake that.  We need to move the financial system to space-time phases in which bubbles are deflated before they grow too big and debt is not allowed to grow to unserviceable levels.  We need to analyse the effects of proposed policies like separating retail and investment banking, introducing a tax on all financial transactions, imposing time delays on trades.  The models need to include the decision-making behaviour of real people and their confidence, not just money.

Social order can be very fragile as seen in the recent riots in England.  Almost certainly this is a system with (at least) two phases: order and anarchy.  We must understand which management strategies make social order more stable.  The models need to include such factors as feeling of belonging and feeling one has a future.  Stability is not the only desirable feature, of course; one must also address the nature of the resulting social order.  Social order is a subject with a long history, e.g.~\cite{HechterHorne}, and forms the core of sociology, yet we believe that the time is ripe for serious advances of a mathematical nature.

Economies are notoriously difficult to manage.  The business cycle and its more extreme versions like recession correspond to long-range correlations in space and time of the space-time phase.  It might be that this is a natural result of seeking to maximise productivity, just as seeking to pass heat faster through a fluid layer leads it to form convection rolls.  Is there some control strategy which can achieve as much productivity without the large scale oscillations?  Is there a control strategy that can achieve it with close to full employment, a goal that would be fulfilling for most people and surely could be more productive?  One of the issues this example raises is that planners often require more than one, often incompatible, objective.  Thus for example, the USA Federal Reserve ``sets the nation's monetary policy to promote the objectives of maximum employment, stable prices, and moderate long-term interest rates'', but admits that ``tensions among the goals can arise in the short run''.  It may be that these goals are incompatible in not just the short run but for ever and, in the terms of Herbert Simon, have to be satisficed \cite{simon}.

Several countries are having immense trouble with the organisation of their health services.  Yet simple measures such as hygiene, prevention and control could substantially shift the space-time phase to one that is much better.

We are struck by the images of famine in Somalia, but drought and malnutrition are a regular feature of the probability distribution there and indeed in other parts of the world too.  Is there a way to manage food production and distribution that would avoid the extreme of hunger and the opposite extremes of obesity in the USA and UK?  The problem is linked to demographics and to social unrest (particularly in the form of civil war).  So another moral emerges here, that it is hard to treat a system in isolation.  Virtually every system has to be considered as open to external influences.  This does not cause a great conceptual shift, but one needs to model the probability distribution for the external influences and if they are themselves the results of space-time phases for large complex systems this is not straightforward.

Epidemiology is a branch of complex systems science in which control is relatively well developed.  Governments have vaccination policies, movement reduction policies, and identification policies for tracking down outbreaks of disease and limiting their spread.  These have been learnt by bitter experience.  Similar ideas apply to the spread of computer viruses.  An area in which there is need for more work is the spread of ideas:  some are deemed good, such as those has leading to reductions in smoking;  while others are considered bad, such as radicalisation (but this depends on the belief system and who you ask).  In either case, it is important to understand what makes an idea spread or not.

Electricity distribution is moving into unknown territory with widely distributed generation, often from highly variable sources like wind, and the consequent problems of balancing supply and demand.  The longterm solution is almost certainly a real-time (and space) pricing signal, coupled to smart consumption, generation and storage devices which take or provide power when it is advantageous to them and not otherwise.  How to design such a pricing system to run stably is a major question.  Stability here does not mean that there would be no fluctuations; it means that the space-time phase would not have any large excursions.

Population, its geographical distribution, age structure and skill distribution is an important issue.  We have now passed seven billion people worldwide and many of the tensions in the world can be attributed to there being too many of us for current technologies to cope with.  Possibly this is a system which has not yet reached a space-time phase but it may be advantageous to manage it onto one.  For example, a simple way to reduce family size is careers for women.  Thus education leading to more women feeling they want a career is probably a good longterm solution.  But to model this in any serious way is a challenge.

Lastly, climate is the archetype of a space-time phase.  Although there are crucial aspects such as cloud formation for which good models are not yet known, one can hope to devise control strategies that move the climate in preferred directions.  The main currently active control is CO$_2$ emissions and various geo-engineering controls have been proposed.

There is clearly an enormous gap between these real world problems and the nascent theory of management of complex systems sketched above.  The biggest challenge is to develop models of social systems that capture the essence of human behaviour.

\section{Conclusion: Complexity Science in FuturICT}

The FuturICT Flagship programme is built on the three pillars of complexity science, social science and ICT. In this paper we have laid out the main concerns and challenges in the science of complex systems with special emphasis on the {\it Complex Systems route to Social Sciences}.

Although there is no agreement on a precise definition of the word `complex', there is wide consensus on the properties that can make systems complex. These include them having: many heterogeneous interacting parts; multiple scales; complicated transition laws; unexpected or unpredicted emergence; sensitive dependence on initial conditions; path-dependent dynamics; networked hierarchical connectivities; interaction of autonomous agents; self organisation; non-equilibrium dynamics; combinatorial explosion; adaptivity to changing environments; co-evolving subsystems; ill-defined boundaries; and multilevel dynamics. In this context, science is seen as the process of abstracting the dynamics of systems from data. This presents many challenges including: data gathering by large-scale experiment, participatory sensing and social computation, and managing huge distributed dynamics and heterogeneous databases; moving from data to dynamical models, going beyond correlations to cause-effect relationships, understanding the relationship between simple and comprehensive models with appropriate choices of variables, ensemble modeling and data assimilation, and modeling systems of systems of systems with many levels between micro and macro; and formulating new approaches to prediction, forecasting, and risk, especially in systems that can reflect on and change their behaviour in response to predictions, and systems whose apparently predictable behaviour is disrupted by apparently unpredictable rare or extreme events.

Undoubtedly great progress is being made, and European scientists are playing a leading role in this field. The ambitions of the FuturICT Flagship Project are high indeed and huge advances in the science of complex systems will be necessary for them to be achieved. ICT will continue to be at the heart of Complexity Science and this science will generate many new ICT applications. Complex systems science desperately needs to be better assimilated with social science and there are enormous challenges and opportunities in this respect. Despite great progress, the science of complex systems is still in its infancy and we must not promise too much.  This makes FuturICT very high risk, but it is hard to see how humankind can face the future without rapid advances in the science of complex systems.\\

\noindent
\textbf{ACKNOWLEDGEMENTS}

\vspace{0.08in}
\noindent
MSM acknowledges financial support for research on Complex Systems from MINECO FIS2007-60327. He also thanks Emilio Hernandez-Garcia for enlightening discussions on the subject.
RSM is grateful to the Alfred P.~Sloan Foundation (New York) for a grant that is enabling him to address the research programme outlined in the section on ``Control and management of complex global systems"
We are all grateful to the Future and Emerging Technology (FET) unit of the European Commission for the support it has given to developing and coordinating complex systems science over the last decade.


\begin{thebibliography}{}

\bibitem{Anderson} P. W. Anderson, Science \textbf{177}, (1972) 393-396

\bibitem{pietronero_2008} L. Pietronero, {\em Complexity ideas from condensed matter and statistical physics}, {\em Europhysics News} {\bf 39},
(2008) 26--29

\bibitem{HelbingComplexity} \textit{Managing Complexity: Insights, Concepts, Applications}, D. Helbig ed. (Springer) 2008

\bibitem{pietronero_2010} L. Pietronero, {\em Physicists get social}, {\em Nature Physics} {\bf 6}, (2010) 641–-640

\bibitem{rmp_2009} C. Castellano, S. Fortunato and V. Loreto, \textit{Statistical physics of social dynamics}, {\em Rev. Mod. Phys.}, {\bf 81} (2009) 591--646

\bibitem{Ball_2004} Ball, P., {\em Critical Mass: How One Thing Leads to Another}, (Farrar, Straus and Giroux, London), 2004

\bibitem{Majorana_1942} Majorana, E., {\em Scientia} {\bf 36}, (1942) 58 and {\em Quant. Finance} {\bf 5}, (2005) 133

\bibitem{Schelling} T. Schelling, \textit{Micromotives and Macrobehavior} (Norton) 1978

\bibitem{Axelrod} R. Axelrod, J. Conflict Resolution \textbf{41}, (1997) 203

\bibitem{Centola} D. Centola, J.C. Gonz{\'a}lez-Avella, V.M. Egu{\'\i}luz and M. San Miguel, J. Conflict Resolution \textbf{51}, (2007) 905

\bibitem{johnson} J. H. Johnson, The future of the social sciences and humanities in the science of complex systems, {\em The European Journal of Social Science Research}, {\bf 23}(2), 2010, 115--134


\bibitem{pluralistic} D.Helbing, Science and Culture,{\bf 76}(2), 2010, 315; arXiv:1007.2818


\bibitem{vicsek_1995} Vicsek T, Czirok A, Ben-Jacob E, Cohen I, Shochet O, {\em Novel type of phase transition in a system of self-driven particles} {\em Phys. Rev. Lett.} {\bf 75}, (1995) 1226

\bibitem{Granger} C. Granger, Econometrica \textbf{37}, (1969) 424-438

\bibitem{Kenett2010} D.Y. Kenett, M. Tumminello, A. Madi, G. Gur-Gershgoren, R. N. Mantegna and E. Ben-Jacob, PLoS ONE 5(12): e15032 (2010)

\bibitem{Hempel2011} S.Hempel, A.Koseska, J.Kurths, and Z. Nikoloski, Phys. Rev. Lett. \textbf{107}, (2011) 054101

\bibitem{wiener} N. Wiener, Cybernetics: Or Control and Communication in the Animal and the Machine. Paris, (Hermann and Cie), Camb. Mass. (MIT Press)

\bibitem{soros} G. Soros, The New Paradigm for Financial Markets. Public Affairs, (2008), New York

\bibitem{Jerusalem} Jerusalem Declaration on Data Access, Use and Dissemination, 2008 http://portale.unipa.it/ocs/home/jerusalemSubscription.html

\bibitem{Science} D. Lazer et al.: Computational Social Science, Science \textbf{323}, (2009) 721-723

\bibitem{HelbingData} D. Helbing and S. Balietti, The European Physical Journal Special Topics \textbf{195}, (2011) 3-68

\bibitem{NATUS} Putting People on the Map: Protecting Confidentiality with Linked Social-Spatial Data, www.nap.edu/catalog/11865.html (Natl Acad. Sci., Washington DC, 2007)

\bibitem{HelbingSimulation} D. Helbing, W. Yu, Proc. Natl. Acad. Sci. (PNAS) USA \textbf{107}, (2010)5265-5266

\bibitem{Araujo} M. B. Araujo and M. New, Trends in Ecology and Evolution \textbf{22}, (2006) 42-47

\bibitem{Bates} J. M. Bates, C. W. Granger, Operat. Res. Quart \textbf{20}, (1969) 451-468

\bibitem{Gozolchiani} A. Gozolchiani, S. Havlin and K. Yamasaki, Phys. Rev. Lett. \textbf{107}, (2011) 148501

\bibitem{Kalnay} E. Kalnay, \textit{Atmospheric modeling, data assimilation, and predictability} (Cambridge University press, 2003)

\bibitem{ws98} D.J. Watts and S. H. Strogatz, Nature \textbf{393}, (1998) 440

\bibitem{ba99} A. L. Barabasi and R. Albert, Science \textbf{286}, (1999) 509

\bibitem{blmch06} S. Bocaletti,V. Latora, Y. Moreno, M. Chavez and D. U. Hwang, Phys. Rep.  \textbf{424}, (2006) 175

\bibitem{bbv08} A. Barrat, M. Barthelemy and A. Vespignani, \textit{Dynamical processes on complex networks} (Cambridge University press, 2008)

\bibitem{cbbgja09} R. Carvalho, L. Buzna, F. Bono, E. Guti\'errez, W. Just and D. Arrowsmith, Phys. Rev. E \textbf{80}, (2009) 016106

\bibitem{bppsh10} S. Buldyrev, R. Parshani, G. Paul, H. E. Stanley and S. Havlim, Nature \textbf{464}, (2010) 1025

\bibitem{herrmann} M. Gonz{\'a}lez, P. Lind and H. Herrmann, Phys. Rev. Lett. \textbf{96}, (2006) 088702

\bibitem{15M} Javier Borge-Holthoefer, Alejandro Rivero, Iñigo García, Elisa Cauhé, Alfredo Ferrer, Dario Ferrer, David Francos, David Iñiguez, María Pilar Pérez, Gonzalo Ruiz, Francisco Sanz, Fermín Serrano, Cristina Viñas, Alfonso Tarancon and Yamir Moreno, PLoS ONE ONE 6(8): e23883 (2011)

\bibitem{messenger} J. Leskovec and E. Horvitz, \textit{Planetary-scale views on a large instant-messaging network} (Proceeding of the 17th international conference on World Wide Web, ACM, 2008)

\bibitem{adkmz08} A. Arenas, A. D{\'\i}az-Guilera, J. Kurths, Y. Moreno adn C. Zhou, Phys. Rep. \textbf{469}, (2008) 93

\bibitem{coev1} M. Zimmermann, V. M. Egu{\'\i}luz and M. San Miguel, Phys. Rev. E \textbf{69}, (2004) 065102

\bibitem{coev2} F. Vazquez, V. M. Egu{\'\i}luz and M. San Miguel, Phys. Rev. Lett. \textbf{100}, (2008) 108702

\bibitem{review_temporal_networks} P. Holme and J. Saram{\"a}ki, ArXiv 1108.1780 (2011), Physics Reports (in press 2012)

\bibitem{Havlin2012} S. Havlin et al., \textit{Challenges of network science: Applications to infrastructures, climate, social systems and economics} in this volume.

\bibitem{Gonzalez-Avella2011} J.C. Gonz{\'a}lez-Avella, V.M. Egu{\'\i}luz, M. Marsili, F. Vega-Redondo and M. San Miguel, \textit{Threshold learning dynamics in social networks}, PLoS ONE 6(5), e20207 (2011)

\bibitem{Johnson2009} P.M. Johnson, Human centered information integration for the smart grid, Technical Report, University of Hawaii, 2009. http://csdl.ics.hawaii.edu/techreports/09-15/09-15.pdf

\bibitem{Gore_2007} Al Gore, {\em An inconvenient truth} (2007)

\bibitem{Randers_2004} Randers J. Meadows D. and Meadows D, {\em Limits to Growth} (Chelsea Green Publishing) 2004

\bibitem{EEA_2009} J. McGlade, {\em Climate crisis needs empowered people}, (2009) \url{http://news.bbc.co.uk/1/hi/sci/tech/7893230.stm}.

\bibitem{barabasi_science_2009} Wang, Pu and Gonz\'{a}lez, Marta C. and Hidalgo, Cesar A. and Barabasi, Albert-Laszlo, {\em Understanding the Spreading Patterns of Mobile Phone Viruses}, {\em Science}, {\bf 324}, (2009) 1071--1076

\bibitem{red_ballon} Pickard, Galen and Pan, Wei and Rahwan, Iyad and Cebrian, Manuel and Crane, Riley and Madan, Anmol and Pentland, Alex, {\em Time-Critical Social Mobilization}, {\em Science}, {\bf 334}, (2011) 509-512

\bibitem{Paulos_2007} E. Paulos, R. Honicky, and E. Goodman, {\em Sensing atmosphere workshop}, Position paper for the sensing on everyday mobile phones in support of participatory research, ACM SenSys, (2007)

\bibitem{EveryAware} EveryAware Project, \url{www.everyaware.eu}

\bibitem{vonahn_2006} Luis von Ahn, {\em Games with a Purpose}, {\em Computer}, {\bf  39}(6), (2006) 92--94

\bibitem{vonahn_2004} Luis von Ahn and Laura Dabbish, {\em Labeling images with a computer game}, CHI '04: Proceedings of the SIGCHI conference on Human factors in computing systems, (2004) 319--326

\bibitem{Chilton_2009} Lydia B. Chilton, Clayton T. Sims, Max Goldman, Greg Little, and Robert C. Miller, {\em Seaweed: a web application for designing economic games}, in Proceedings of the ACM SIGKDD Workshop on Human Computation, HCOMP '09, pages 34--35, New York, NY, USA, (2009)

\bibitem{Paolacci_2010} Gabriele Paolacci, Chandler Jesse, and Panagiotis G. Ipeirotis, {\em Running experiments on amazon mechanical turk} {\em Judgment and Decision Making} {\bf 5} (2010) 411--419

\bibitem{Huberman_2011} Zeinab Abbassi, Christina Aperjis, Bernardo A. Huberman, {\em Swayed by Friends or by the Crowd?}, \url{http://arxiv.org/abs/1111.0307} (2011)

\bibitem{Suri_2010} Siddharth Suri and Duncan J. Watts, {\em Cooperation and Contagion in Networked Public Goods Experiments} {\em Computing Research Repository (CORR)} 2010

\bibitem{GilesNature} J. Giles, Nature \textbf{470}, (2011) 18-19

\bibitem{WardPNAS} A. J. W. Ward, et al., Proc. Nat. Acad. Sci. (PNAS) USA \textbf{108}, (2011) 2312-2315

\bibitem{Conradt} L. Conradt, Nature \textbf{471}, (2011) 40-41

\bibitem{Crutchfield2010} J. P. Crutchfield, W. L. Ditto and S. Sinha, Chaos \textbf{20}, (2010) 037101(1-6)

\bibitem{Jaeger} H. Jaeger and H. Haas, Science \textbf{304}, (2004) 78-80

\bibitem{Bunomano} D. V. Buonomano and W. Maass, Nat. Rev. Neurosc. \textbf{10}, (2009) 113

\bibitem{Appeltant} L. Appeltant, M.C. Soriano, G. Van der Sande, J. Danckaert, S. Massar, J. Dambre, B. Schrauwen, C.R. Mirasso, and I. Fischer, Nature Communications \textbf{2}, (2011) 468

\bibitem{Trist1951} E.L. Trist and K.W. Bamforth, {\em Some social and psychological consequences of the longwall method of coal getting: An Examination of the Psychological Situation and Defences of a Work Group in Relation to the Social Structure and Technological Content of the Work System}, Human Relations \textbf{4}, (1951) 3

\bibitem{Grabowicz2012} P. Grabowicz, J. J. Ramasco, E. Moro, J. M. Pujol, V. M. Egu{\'\i}luz, PLoS ONE 7(1), e29358 (2012)

\bibitem{Sornette2003} Sornette D (2003): \textit{Why Stock Markets Crash: Critical Events in Complex Financial Systems} (Princeton Univ. Press.)

\bibitem{Albeverio2006} Albeverio S., V. Jentsch and H. Kantz (eds) (2006); Extreme Events in Nature and Society, Springer, Berlin/Heidelberg/New York

\bibitem{Erdi} P. \'Erdi, \textit{Complexity Explained} (Springer Verlag, Berlin-Heidelberg, 2007)

\bibitem{bak96} P. Bak, \textit{How Nature Works: The Science of Self-Organized Criticality}, (New York: Copernicus, 2007)

\bibitem{baketal87} P. Bak, C. Tang and K. Wiesenfeld, Physical Review Letters \textbf{59} (1987) 381-384

\bibitem{Sornette2002} D. Sornette, Predictability of catastrophic events: Material rupture, earthquakes, turbulence, financial crashes, and human birth, Proc. Natl. Acad. Sci. (PNAS) USA 2002

\bibitem{Sontag} E.D.Sontag, Mathematical control theory (Springer, 2nd ed, 1998)

\bibitem{Mac08} RS MacKay, Nonlinearity in Complexity Science, Nonlinearity 21 (2008) T273--281

\bibitem{Mac11} RS MacKay, Robustness of Markov processes on large networks, J Difference Eqns \& Applns 17 (2011) 1155--67

\bibitem{DM11} M Diakonova, RS MacKay, {\em Mathematical examples of space-time phases}, {\em Int J Bif Chaos} 21 (2011) 2297--2304

\bibitem{Liggett} T.M.Liggett, Interacting Particle Systems (Springer, reprinted 2004)

\bibitem{HechterHorne} M.Hechter, C.Horne, Theories of social order (Stanford Univ Press, 2003)

\bibitem{simon} H. Simon, {\em The Sciences of the artificial}, (MIT Press, Cambridge, Mass), 1969


\bibitem{sornier} D. Sornier, Dragon Kings, Black Swans and the Prediction of Crises. International Journal of Terraspace Science and Engineering, 2009. http://ssrn.com/abstract=1470006

\bibitem{helbing2000} D. Helbing, I. Farkas, T. Vicsek, Simulating dynamical features of escape panic.
    Nature 407, 487-490 (28 September 2000) | doi:10.1038/35035023

\bibitem{helbing2007}D. Helbing, A. Johansson, H. Z. Al-Abideen (2007) The Dynamics of Crowd Disasters: An Empirical Study. Phys. Rev. E 75, 046109

\end{thebibliography}
\end{document}